\documentclass[11pt]{article}
\usepackage[margin=1in]{geometry}
\usepackage{graphicx}

\usepackage{lipsum}
\usepackage{epsfig}
\usepackage{amsmath}
\usepackage{amssymb}
\usepackage{amsfonts}
\usepackage{multicol}
\usepackage{graphicx,url}
\usepackage[english]{babel}   
\usepackage[utf8]{inputenc}
\usepackage{xfrac}
\usepackage{multirow}
\usepackage{geometry}
\usepackage{bbm}

\def\beq{\begin{equation}}
\def\eeq#1{\label{#1}\end{equation}}
\def\eeqn{\end{equation}}

\def\beqa{\begin{eqnarray}}
\def\eeqa#1{\label{#1}\end{eqnarray}}
\def\eeqan{\end{eqnarray}}

\let\bar=\overbar

\def\Dslash{\not{\hbox{\kern-4pt $D$}}}
\def\dslash{\not{\hbox{\kern-2pt $\del$}}}

\def\msb{{\bar{\ssstyle M \kern -1pt S}}}

\def\Title#1{\begin{center} {\Large {\bf #1} } \end{center}}
\def\Author#1{\begin{center} {\normalsize {\sc #1} } \end{center}}
\def\Institution#1{\begin{center} {\normalsize {\it #1} } \end{center}}
\def\Abstract#1{\noindent {\normalsize {\bf Abstract:} {\normalfont #1}}}
\def\Conference{\vspace{4mm}\begin{raggedright} {\normalsize {\it Talk presented at the 2019 Meeting of the Division of Particles and Fields of the American Physical Society (DPF2019), July 29--August 2, 2019, Northeastern University, Boston, C1907293.} } \end{raggedright}\vspace{4mm}}

\begin{document}

%
%

\Title{A $U(1)_X$ extension to the MSSM with three families}

\Author{J.S. Alvarado, Carlos E. Diaz, R. Martinez. }

\Institution{Departamento de F\'{i}sica$,$ Universidad Nacional de Colombia\\
 Ciudad Universitaria$,$ K. 45 No. 26-85$,$ Bogot\'a D.C.$,$ Colombia}

\Abstract{
We propose a supersymmetric extension of the anomaly-free and three families nonuniversal $U(1)$ model, with the inclusion of four Higgs doublets and four Higgs singlets. The quark sector is extended by adding three exotic quark singlets, while the lepton sector includes two exotic charged lepton singlets, three right-handed neutrinos and three sterile Majorana neutrinos to obtain the fermionic mass spectrum. By implementing an additional $\mathbb{Z}_2$ symmetry, the Yukawa coupling terms are suited in such a way that the fermion mass hierarchy is obtained without fine-tuning. The effective mass matrix for SM neutrinos is fitted to current neutrino oscillation data to check the consistency of the model with experimental evidence, obtaining that the normal-ordering scheme is preferred over the inverse ones. The electron and up, down and strange quarks are massless at tree level, but they get masses through radiative correction at one loop level coming from the sleptons and Higgsinos contributions. We show that the model predicts a like-Higgs SM mass at electroweak scale by using the VEV according to the symmetry breaking and fermion masses.}

\Conference

%
%

\section{Introduction}
A $U(1)_{X}$ extension of the Minimal Supersymmetric Standard Model is built and explored by considering a non-universal X charge assignation that makes the model anomaly free with the inclusion of additional quarks, leptons, Higgs doublets and singlets \cite{PRD}. The standard model fermion sector is examined and provides neutrino masses through a see saw mechanism ensuring the correct squared mass differences and the correct CKM and PMNS mixing matrices because a previous work result. Furthermore, 3 fermions become massless but acquire a finite mass value through radiative corrections thanks to the coupling to a scalar Higgs singlet which does not acquire a VEV. The scalar sector is explored and it is found the conditions for having the lightest of them as a 125 GeV  CP-even Higgs boson  which depends only on the GeV VEV. Meanwhile the other mass eigenstates lies in the Tev Scale; as well for the charged and CP-odd higgs bosons. 

\section{Scalar sector}

\begin{table}
\caption{Scalar content of the model, non-universal $X$ quantum number, $\mathbb{Z}_{2}$ parity and hypercharge}
\label{tab:Bosonic-content-A-B}
\centering
\begin{tabular}{lll cll}\hline\hline 
\multirow[l]{3}{*}{
\begin{tabular}{l}
    Higgs    \\
    Scalar  \\
    Doublets
\end{tabular}
}
&\multicolumn{2}{l}{}&
\multirow[l]{3}{*}{
\begin{tabular}{l}
    Higgs    \\
    Scalar  \\
    Singlets
\end{tabular}
}
&\multicolumn{2}{l}{}\\ 
 &&&
 && \\ 
 &$X^{\pm}$&$Y$&
 &$X^{\pm}$&$Y$
\\ \hline\hline 
$\small{\hat{\Phi}_{1}=\begin{pmatrix}\hat{\phi}_{1}^{+}\\\frac{\hat{h}_{1}+v_{1}+i\hat{\eta}_{1}}{\sqrt{2}}\end{pmatrix}}$&$\sfrac{+2}{3}^{+}$&$+1$&
$\hat{\chi}=\frac{\hat{\xi}_{\chi}+v_{\chi}+i\hat{\zeta}_{\chi}}{\sqrt{2}}$	&	$\sfrac{+1}{3}^{+}$	&	$0$	\\
$\small{\hat{\Phi}_{2}=\begin{pmatrix}\hat{\phi}_{2}^{+}\\\frac{\hat{h}_{2}+v_{2}+i\hat{\eta}_{2}}{\sqrt{2}}\end{pmatrix}}$&$\sfrac{+1}{3}^{-}$&$+1$&$\hat{\sigma}$
 &$\sfrac{-1}{3}^{-}$
&	$0$		\\
$\small{\hat{\Phi}^\prime_{1}=\begin{pmatrix}\frac{\hat{h}_{1}'+v_{1}'+i\hat{\eta}_{1}'}{\sqrt{2}}\\\hat{\phi}_{1}^{-\prime}\end{pmatrix}}$&$\sfrac{-2}{3}^{+}$&$-1$&
$\hat{\chi}'=\frac{\hat{\xi}'_{\chi}+v_{\chi}'+i\hat{\zeta}'_{\chi}}{\sqrt{2}}$
	&	$\sfrac{-1}{3}^{+}$
&	0\\
$\small{\hat{\Phi}^\prime_{2}=\begin{pmatrix}\frac{\hat{h}_{2}'+v_{2}'+i\hat{\eta}_{2}'}{\sqrt{2}}\\\hat{\phi}_{2}^{-\prime}\end{pmatrix}}$&$\sfrac{-1}{3}^{-}$&$-1$&$\hat{\sigma}'$
	&$\sfrac{+1}{3}^{-}$
&$0$	\\\hline\hline
\end{tabular}
\end{table}

The Lagrangian for the scalar sector that describes the minimal supersymetric extension to the $U(1)_{X}$ model in the literature is given by the addition of F-terms, D-terms and a soft-supersymetry breaking potential. The F-terms were obtained from the following superpotential:
\begin{align}
    W_{\phi}&=-\mu_{1}\hat{\Phi}'_{1}\hat{\Phi}_{1}-\mu_{2}\hat{\Phi}'_{2}\hat{\Phi}_{2} - \mu_{\chi}\hat{\chi} '\hat{\chi}- \mu_{\sigma}\hat{\sigma} '\hat{\sigma}. 
\end{align}
Therefore, the F-terms' potential for scalar fields is given by
\begin{small}
\begin{align}
-V_{F}&=\mu_{1}^2(\Phi_{1}^\dagger\Phi_{1}+\Phi_{1}^{\prime\dagger}\Phi_{1}^{\prime})+\mu_{2}^2(\Phi_{2}^\dagger\Phi_{2}+\Phi_{2}^{\prime\dagger}\Phi_{2}^{\prime})+\mu_{\chi}^2(\chi^*\chi+\chi^{\prime*}\chi')+\mu_{\sigma}^2(\sigma^*\sigma+\sigma^{\prime*}\sigma').
\end{align}
\end{small}
On the other hand, the D-terms potential, consequence of gauge invariance, reads:

\begin{align}
-V_{D}&=\frac{g^{2}}{2}\Big[ |\Phi_{1}^{\dagger}\Phi_{2}|^{2}+|\Phi_{1}^{\prime\dagger}\Phi_{2}'|^2+|\Phi_{1}^{\prime\dagger}\Phi_{1}|^2+|\Phi_{1}^{\prime\dagger}\Phi_{2}|^2+|\Phi_{2}^{\prime\dagger}\Phi_{1}|^2\nonumber +|\Phi_{2}^{\prime\dagger}\Phi_{2}|^2-|\Phi_{1}|^{2}|\Phi_{2}|^{2} \nonumber\\
&-|\Phi_{1}^{\prime}|^{2}|\Phi_{2}^{\prime}|^{2} \Big] +\frac{g^{2} + g^{\prime 2}}{8}(\Phi_{1}^{\dagger}\Phi_{1}+\Phi_{2}^{\dagger}\Phi_{2}-\Phi_{1}^{\prime\dagger}\Phi_{1}^{\prime}-\Phi_{2}^{\prime\dagger}\Phi_{2}^{\prime})^{2} \nonumber \\
 &+\frac{g_{X}^{2}}{2}\Big[\frac{2}{3}(\Phi_{1}^{\dagger}\Phi_{1}-\Phi_{1}^{\prime\dagger}\Phi_{1}^{\prime})+\frac{1}{3}(\Phi_{2}^{\dagger}\Phi_{2}-\Phi_{2}^{\prime\dagger}\Phi_{2}^{\prime})-\frac{1}{3}(\chi^{*}\chi-\chi^{\prime*}\chi^{\prime})-\frac{1}{3}(\sigma^{*}\sigma-\sigma^{\prime*}\sigma^{\prime})\Big]^{2} 
\end{align}

Finally, the soft supersymmetry breaking Lagrangian turns out to be:

\begin{align}
    -V_{soft}&=m_{1}^{2}\Phi_{1}^{\dagger}\Phi_{1} + {m}_{1}^{\prime 2}{\Phi}_{ 1}^{\prime \dagger}\Phi'_{1} + m_{2}^{2}\Phi_{2}^{\dagger}\Phi_{2} + {m}_{2}^{\prime 2}\Phi _{2}^{\prime\dagger}\Phi'_{2}+m_{\chi}^{2}\chi^{\dagger}\chi + {m}_{\chi}^{\prime 2}{\chi}^{\prime\dagger}\chi' + m_{\sigma}^{2}\sigma^{\dagger}\sigma\nonumber\\
    &+{m}_{\sigma}^{\prime 2}{\sigma}^{\prime\dagger}\sigma' -\mu_{11}^{2}\epsilon_{ij}({\Phi}_{1}^{\prime i}\Phi_{1}^{j}) -\mu_{22}^{2}\epsilon_{ij}({\Phi}_{2}^{\prime i}\Phi_{2}^{j}) -\mu_{\chi\chi}^{2}(\chi\chi') \nonumber\\
    &+ \frac{2\sqrt{2}}{9}(-k_{1}\Phi_{1}^{\dagger}\Phi_{2}\chi' +k_{2}\Phi_{1}^{\dagger}\Phi_{2}\chi^*-k_{3}\Phi_{1}'{}^{\dagger}\Phi_{2}'\chi +k_{4}\Phi_{1}'{}^{\dagger}\Phi_{2}'\chi'{}^*) + h.c.
\end{align}

\subsection{CP-Even scalar particles}
After symmetry breaking all Higgs doublets and singlets aquire a VEV which allows us to construct the following a squared mass matrix for CP-Even scalar particles

\begin{align}\label{h}
    \frac{1}{2}M_{h}^{2}&=\begin{pmatrix}
    M_{\phi} & M_{\phi \chi}  \\
    M_{\phi \chi}^{T} & M_{\chi \chi}  
    \end{pmatrix},
\end{align}

\begin{align*}
M_{\phi}&=\\
&\begin{pmatrix}
    f_{4g}v_{1}^{2}-\frac{v_{2}f_{1k}}{9v_{1}}-\frac{v_{1}'\mu_{11}^{2}}{2v_{1}} & -f_{4g}v_{1}v_{1}'-\frac{\mu_{11}^{2}}{2} &  f_{2g}v_{1}v_{2}+\frac{f_{1k}}{9}&-f_{2g}v_{1}v_{2}'\\
    * & f_{4g}v_{1}'{}^{2}-\frac{v_{2}'f_{2k}}{9v_{1}'}+\frac{v_{1}\mu_{11}^{2}}{2v_{1}'}& -f_{2g}v_{1}'v_{2} &f_{2g}v_{1}'v_{2}'+\frac{f_{2k}}{9}\\
    *&*& f_{1g}v_{2}^{2}-\frac{v_{1}f_{1k}}{9v_{2}}-\frac{v_{2}'\mu_{22}^{2}}{2v_{2}}& -f_{1g}v_{2}v_{2}'-\frac{\mu_{22}^{2}}{2}\\
    *&*&*&f_{1g}v_{2}'{}^{2}-\frac{v_{1}'f_{2k}}{9v_{2}'}-\frac{v_{2}\mu_{22}^{2}}{2v_{2}'},
    \end{pmatrix}    
\end{align*}

\begin{align}\label{11}
   M_{\chi \chi}& =\begin{pmatrix}
   \frac{g_{X}^2}{18}v_{\chi}^2 +\frac{v_{\chi}'\mu_{\chi\chi}^2}{2v_{\chi}}-\frac{k_{2}v_{1}v_{2}-k_{3}v_{1}'v_{2}'}{9v_{\chi}} & -\frac{g_{X}^2}{18}v_{\chi}v_{\chi}'-\frac{\mu_{\chi\chi}^{2}}{2} \\
    * &  \frac{g_{X}^2}{18}v_{\chi}'{}^2 +\frac{v_{\chi}\mu_{\chi\chi}^2}{2v_{\chi}'}-\frac{-k_{1}v_{1}v_{2}+k_{4}v_{1}'v_{2}'}{9v_{\chi}'}
    \end{pmatrix}
\end{align}

\begin{align}
     M_{\phi \chi}&=\frac{1}{9}\begin{pmatrix}
    k_{2}v_{2}-g_{X}^2v_{1}v_{\chi} & -k_{1}v_{2}+g_{X}^2v_{1}v_{\chi}'  \\
    -k_{3}v_{2}'+g_{X}^2v_{1}'v_{\chi} & k_{4}v_{2}'-g_{X}^2v_{1}'v_{\chi}'\\
    k_{2}v_{1}-\frac{1}{2}g_{X}^2v_{2}v_{\chi}& -k_{1}v_{1}+\frac{1}{2}g_{X}^2v_{2}v_{\chi}'\\
    -k_{3}v_{1}'+\frac{1}{2}g_{X}^2v_{2}'v_{\chi}& k_{4}v_{1}'-\frac{1}{2}g_{X}^2v_{2}'v_{\chi}'
    \end{pmatrix}
\end{align}

Where we have defined $f_{ig}=\frac{g^{2}+g'^{2}}{8}+\frac{i}{18}g_{X}^{2}$
The above structure fulfill the conditions for implementing a  seesaw mechanism due to the large value of $\mu_{\chi\chi}$, comming from the soft breaking potential. Then after block diagonalization the mass eigenstates are found in a tree level aproximation.

\begin{align}
  m_{h6}^{2}&\approx\mu_{\chi\chi}^{2}\frac{v_{\chi}^{2}+v_{\chi}'{}^{2}}{v_{\chi}v_{\chi}'}\nonumber \\
    m_{h5}^{2}&\approx\frac{g_{X}^{2}}{9}(v_{\chi}^{2}+v_{\chi}'{}^{2})-\frac{2}{9}\frac{v_{1}v_{2}(k_{2}v_{\chi}'-k_{1}v_{\chi})+v_{1}'v_{2}'(k_{4}v_{\chi}-k_{3}v_{\chi}')}{v_{\chi}v_{\chi}'}\nonumber \\
   m_{h4}^{2}&\approx \mu_{22}^{2}\frac{v_{2}^{2}+v_{2}^{\prime 2}}{v_{2}v_{2}^{\prime}} \nonumber\\
   m_{h3}^{2}&\approx \mu_{11}^{2}\frac{v_{1}^{2}+v_{1}^{\prime 2}}{v_{1}v_{1}^{\prime}}  \nonumber\\
m_{h2}^{2}&\approx \frac{2v^{2}(v_{1}v_{2}(k_{1}v_{\chi}^{\prime}-k_{2}v_{\chi})+v_{1}^{\prime}v_{2}^{\prime}(k_{3}v_{\chi}-k_{4}v_{\chi}^{\prime}))}{9(v_{1}^{2}+v_{1}^{\prime 2})(v_{2}^{2}+v_{2}^{\prime 2})} \nonumber\\
m_{h1}^{2}&\approx \frac{g_{X}^{2}(2v_{1}^{2}+v_{2}^{2}-2v_{1}^{\prime 2}-v_{2}^{\prime 2})^{2} }{9(v_{1}^{2}+v_{2}^{2}+v_{1}^{\prime 2}+v_{2}^{\prime 2})}+ \frac{ (g^{2}+g'{}^{2})(v_{1}^{2}+v_{2}^{2}-v_{1}^{\prime 2}-v_{2}^{\prime 2})^{2}}{4(v_{1}^{2}+v_{2}^{2}+v_{1}^{\prime 2}+v_{2}^{\prime 2})} 
\end{align}

Then, it is possible to get a $125 GeV$ CP-even scalar particle, compatible with the observed Higgs boson for several sets of VEV which should accomplish $v_{1}^{2}+v_{2}^{2}+v_{1}^{\prime 2}+v_{2}^{\prime 2}=v^{2} = (246.22 GeV)^{2}$. This condition is mandatory in order to get the correct value for gauge bosons. For example considering $v'_{1}=20.577$, $v'_{2}=52.577$, $v_{1}^{\prime}=195.681$, $v_{2}^{\prime}=138.367$, and $g_{X}=0.769$ a $125$ GeV is found. Nevertheless, the matrix structire for CP-odd scalar particles and Charged scalar particles shows a very similar one, whose mass eigenstates result in similar expressions.\newline

\subsection{Gauge boson masses}
As consequence of the inclusion of the symmetry $U(1)_{X}$, a new gauge boson $Z'$ is predicted, and it has to be included in the covariant derivate.
\begin{align}
    D_{\mu}=\partial_{\mu} -igW_{\mu}^{a}T_{a} - ig'\frac{Y}{2}B_{\mu}-ig_{X}Z'_{\mu}
\end{align}
Therefore the gauge boson masses are determined by the interaction terms, which are present in the scalar field kinetic terms. On one hand the charged bosons $W^{\pm}_{\mu}=(W_{\mu}^{1}\mp W_{\mu}^{2})/\sqrt{2}$ acquire masses $M_{W}=\frac{gv}{2}$. On the other hand the neutral gauge bosons $(W_{\mu}^{3},B_{\mu},Z'_{\mu})$ make up a squared-mass matrix after SSB given by:
\begin{align}
    M_{0}^{2}=\frac{1}{4}\begin{pmatrix}
    g^{2} v^{2} & -gg'v^{2} & -\frac{2}{3}g g_{X} v^{2}(1+\cos^{2}\beta) \\
    * & g'{}^{2} v^{2} & \frac{2}{3}g'g_{X} v^{2}(1+\cos^{2}\beta) \\
    * & * & \frac{4}{9}g_{X}^{2} V_{\chi}^{2}\left[1+(1+3\cos^{2}\beta)\frac{v^{2}}{V_{\chi}^{2}}\right],
    \end{pmatrix}, \nonumber
\end{align}
where we have defined:
\begin{align}
    v^{2}&=v_{1}^{2}+v_{2}^{2}+{v}_{1}^{\prime 2}+{v}_{2}^{\prime 2} = (246.22 GeV)^2 \label{v}\\
    \tan{\beta}&=\frac{\sqrt{v_{2}^{2}+{v}_{2}^{\prime 2}}}{\sqrt{v_{1}^{2}+{v}_{1}^{\prime 2}}} \equiv \frac{V_{2}}{V_{1}}\\
    V_{\chi}^{2}&\equiv v_{\chi}^{2} + {v}_{\chi}^{\prime 2}\label{vx}
\end{align}
Despite in this model we have 4 Higgs doubles, it recreates the same mass structure found in \cite{no-susy} when adopting the definitions (\ref{v}-\ref{vx}). Thus, in the gauge boson sector we can split in two the acceptable parameter values given in the same reference in order to know the possible values for doublets VEV. Nevertheless, it also means that the neutral boson mass eigenvalues are already determined.
\begin{align}
    M_{\gamma}&=0 & M_{Z}&\approx\frac{gv}{2\cos{\theta_{W}}}  &    M_{Z'}&\approx \frac{g_{X}V_{\chi}}{3},
\end{align}
where $\tan\theta_{W}=\frac{g'}{g}$, as it is defined in the standard model. Also, the PMNS matrices \cite{4}, \cite{5} are reproduced since in the no-SUSY version of the theory guaranties it.

\section{Fermion sector}
\subsection{Quark masses}
\begin{table}
    \centering
    \caption{Quark content of the abelian extension, non-universal $X$ quantum number and parity $\mathbb{Z}_{2}$.}
\begin{tabular}{lll lll}\hline\hline 
\multirow[l]{3}{*}{
\begin{tabular}{l}
    Left-    \\
    Handed  \\
    Fermions
\end{tabular}
}
&\multicolumn{2}{l}{}&
\multirow[l]{3}{*}{
\begin{tabular}{l}
    Right-    \\
    Handed  \\
    Fermions
\end{tabular}
}
&\multicolumn{2}{l}{}\\ 
 &&&
 && \\ 
 &$X^{\pm}$&&
 &$X^{\pm}$&
\\ \hline\hline 
\multicolumn{6}{c}{SM Quarks}\\ \hline\hline	
\begin{tabular}{c}	
	$ \hat{q} ^{1}_{L}=\begin{pmatrix}\hat{u}^{1}	\\ \hat{d}^{1} \end{pmatrix}_{L}$  \\
	$  \hat{q} ^{2}_{L}=\begin{pmatrix}\hat{u}^{2}	\\ \hat{d}^{2} \end{pmatrix}_{L}$  \\
	$  \hat{q} ^{3}_{L}=\begin{pmatrix}\hat{u}^{3}	\\ \hat{d}^{3} \end{pmatrix}_{L}$ 
\end{tabular} &
\begin{tabular}{c}
		$\sfrac{+1}{3}^{+}$	\\
	\\	$0^{-}$	\\
	\\	$0^{+}$	\\
\end{tabular}   &
\begin{tabular}{c}
			\\
	\\		\\
	\\		\\
\end{tabular}   &
\begin{tabular}{c}
	$ \begin{matrix}\hat{u}^{1\; c }_{L}	\\ \hat{u}^{2\; c}_{L} \end{matrix}$  \\
	$ \begin{matrix}\hat{u}^{3\; c}_{L}	\\ \hat{d}^{1\; c }_{L} \end{matrix}$  \\
	$ \begin{matrix}\hat{d}^{2\; c }_{L}	\\ \hat{d}^{3\; c }_{L} \end{matrix}$ 
\end{tabular} &
\begin{tabular}{c}
	$ \begin{matrix} \sfrac{-2}{3}^{+}	\\ \sfrac{-2}{3}^{-} \end{matrix}$  \\
	$ \begin{matrix} \sfrac{-2}{3}^{+}	\\ \sfrac{+1}{3}^{-} \end{matrix}$  \\
	$ \begin{matrix} \sfrac{+1}{3}^{-}	\\ \sfrac{+1}{3}^{-} \end{matrix}$ 
\end{tabular} &
\begin{tabular}{c}
	\\
	 \\
\end{tabular}
\\ \hline\hline 

\multicolumn{6}{c}{Non-SM Quarks}\\ \hline\hline	
\begin{tabular}{c}	
	$\hat{\mathcal{T}}_{L}$	\\		\\
	$\mathcal{J}_{L}^{1}$	\\	$\mathcal{J}_{L}^{2}$	
\end{tabular} &
\begin{tabular}{c}
	$\sfrac{+1}{3}^{-} $	\\	\\	
	$ 0^{+} $           	\\	$ 0^{+} $
\end{tabular}   &
\begin{tabular}{c}
		\\		\\	
		\\		
\end{tabular}   &

\begin{tabular}{c}
	$\hat{\mathcal{T}}_{L}^{c}$	\\		\\
	$\hat{\mathcal{J}}_{L}^{c}$	\\	$\hat{\mathcal{J}}_{L}^{c \ 2}$	
\end{tabular} &
\begin{tabular}{c}
	$\sfrac{-2}{3}^{-} $	\\	              	\\	
	$\sfrac{-1}{3}^{+} $	\\	$\sfrac{-1}{3}^{+} $
\end{tabular} &
\begin{tabular}{c}
		\\		\\	
		\\		
\end{tabular}
\\ \hline\hline 

\end{tabular}
\end{table}

According to the $SU(2)_{L}\otimes U(1)_{Y}\otimes U(1)_{X}\otimes Z_{2}$ symmetry, the most general superpotencial for the quarks superfields is given by:\\
\begin{small}
\begin{align}
    W_{Q}&=\hat{q}_{L}^{1}\hat{\phi}_{2}h_{2u}^{12}(\hat{U}_{L}^{2})^{C} + \hat{q}_{L}^{2}\hat{\phi}_{1}h_{1u}^{22}(\hat{U}_{L}^{2})^{C} + \hat{q}_{L}^{3}\hat{\phi}_{1}h_{1u}^{3k}(\hat{U}_{L}^{k})^{C} - \hat{q}_{L}^{3}\hat{\phi '}_{2}h_{2d}^{3j}(\hat{D}_{L}^{2})^{C}+ \hat{q}_{L}^{1}\hat{\phi}_{2}h_{2T}^{1}\hat{T}_{L}^{C} \nonumber\\
    &+\hat{q}_{L}^{2}\hat{\phi}_{1}h_{1T}^{2}\hat{T}_{L}^{C}-\hat{q}_{L}^{1}\hat{\phi '}_{1}{h}_{1J}^{1a}(\hat{J}_{L}^{a})^{C} - \hat{q}_{L}^{2}\hat{\phi '}_{2}{h}_{2J}^{2a}(\hat{J}_{L}^{a})^{C} + \hat{T}_{L}\hat{\chi}'g_{\chi' T}\hat{T}_{L}^{C} \nonumber\\
    &+ \hat{J}_{L}^{a}\hat{\chi}g_{\chi J}^{ab}(\hat{J}_{L}^{b})^{C} +\hat{T}_{L}\hat{\chi '}h_{\chi' u}^2(\hat{U}_{L}^{2})^{C} +  \hat{J}_{L}^{a}\hat{\sigma}h_{\sigma J}^{aj}(\hat{D}_{L}^{j})^{C}+\hat{T}_{L}\hat{\sigma '}{h}_{\sigma' T}^{k}(\hat{U}_{L}^{k})^{C}, 
\end{align}
\end{small}
It allows us to construct the squared mass matrices for up-like and down-like quarks. In both cases it is found that SM model particles  mixing matrix depends on the small valued VEV $v_{i}, v_{i}'$, $i=1,2$ while the mixing matrix between exotic particles turns out to depend on the singlet VEV $v_{\chi}, v_{\chi}'$. In this way a seesaw mechanism can be implemented, in such a way that the rotated mass matrices can be read as follows.

\begin{align*}
\mathbbm{m}_{U}^{2}&=\left(V_{L}^{(U)}\right)^{T} \mathbb{M}_{U}^{2} V_{L}^{(U)}=\begin{pmatrix}
m_{U}^{2} & 0 \\
0 & m_{T}^{2}
\end{pmatrix} 
& \mathbbm{m} _D^2&=\left(V_{L}^{(D)}\right)^T \mathbb{M}_{D}^{2} V_{L}^{(D)}=\begin{pmatrix}
m_D^2 & 0 \\
0 & m_J^2
\end{pmatrix}, \\
m_U^2 &\approx  \frac{1}{2}\begin{pmatrix}
\upsilon _2^2  r_1^2 & \upsilon _1 \upsilon _2  r_1r_2 & 0 \\
\upsilon _1 \upsilon _2  r_1r_2 & \upsilon _1^2  r_2^2 & 0 \\
0 & 0 & \upsilon _1 ^2 \left(a_{31}^2+a_{33}^2 \right)
\end{pmatrix} & m_D^2 &=\frac{1}{2}\begin{pmatrix}
0 & 0 &  0    \\
0 & 0 & 0  \\
0 & 0 &  \upsilon _2 ^2 \left(B_{31}^2+B_{32}^2+B_{33}^2 \right) 
\end{pmatrix}, \\
m_{T}^{2} & \approx \frac{1}{2}\upsilon _{\chi}^2\left(c_2^2+h_{\chi }^{T2}\right) & m_{J}^{2} & \approx  \frac{ \upsilon _{\chi }^2}{2}\begin{pmatrix}
k_{11}^2 & 0  \\
 0 &   k_{22}^2
\end{pmatrix}.
\end{align*}

\begin{figure}
\centering
\includegraphics[scale=0.16]{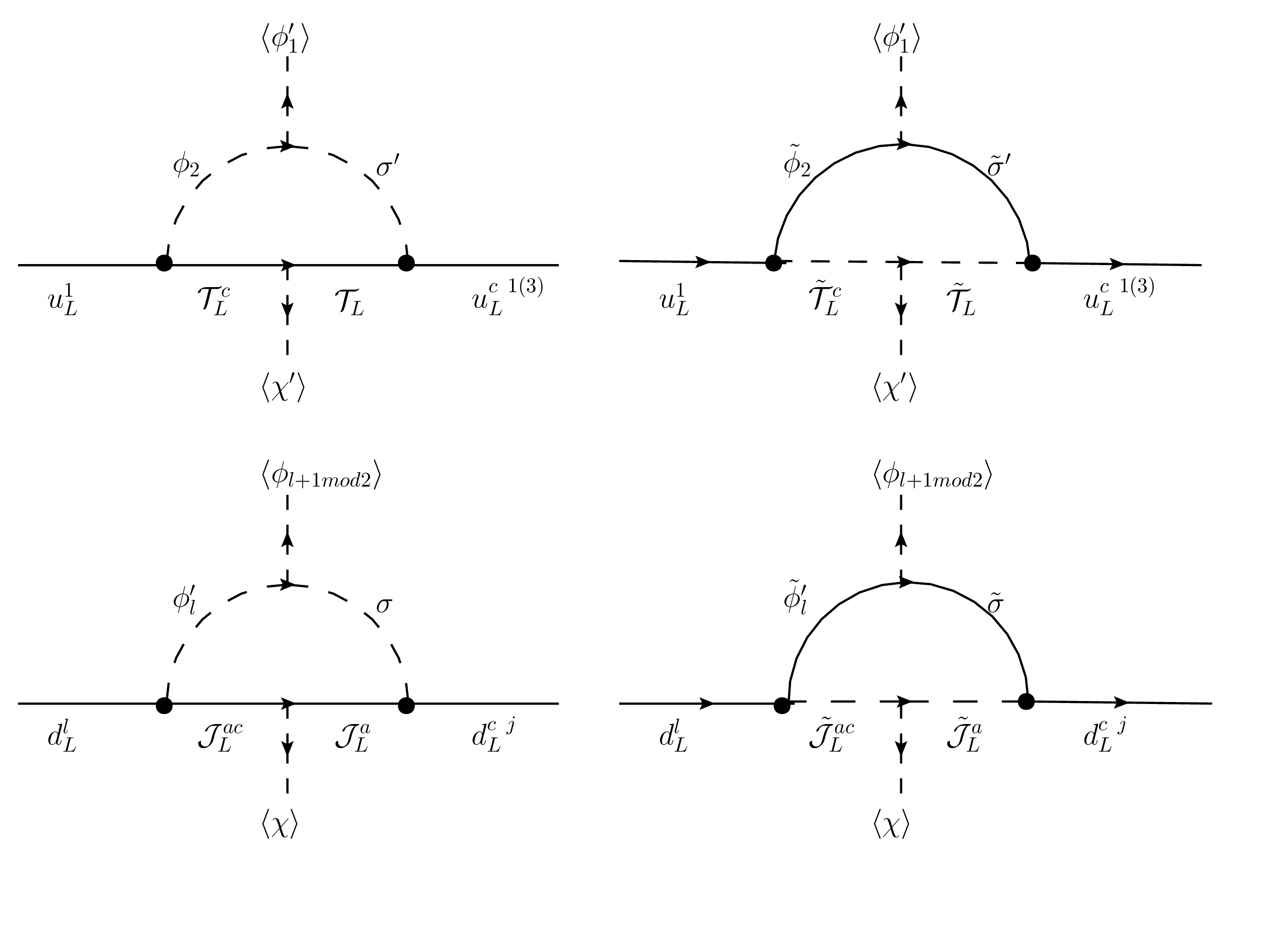}
\caption{Mass one-loop correction for (a) up and (b) down
sector, where $k, l, m, n =1, 2$ and $j = 1, 2, 3$.}
\label{loopquark}
\end{figure}

Despite the standard model particles lies on the electroweak scale, this mass matrices gives as a result a massless up, down and strange quark. Nonetheless they can aquire a mass value through radiative corrections thanks to the coupling to the scalar singlet $\sigma$ as shown in figure \ref{loopquark}

\subsection{Leptons}
\begin{table}
    \centering
    \caption{Lepton content of the abelian extension, non-universal $X$ quantum number and parity $\mathbb{Z}_{2}$.}
\begin{tabular}{lll lll}\hline\hline 
\multirow[l]{3}{*}{
\begin{tabular}{l}
    Left-    \\
    Handed  \\
    Fermions
\end{tabular}
}
&\multicolumn{2}{l}{}&
\multirow[l]{3}{*}{
\begin{tabular}{l}
    Right-    \\
    Handed  \\
    Fermions
\end{tabular}
}
&\multicolumn{2}{l}{}\\ 
 &&&
 && \\ 
 &$X^{\pm}$&&
 &$X^{\pm}$&
\\ \hline\hline 
\multicolumn{6}{c}{SM Leptons}\\ \hline\hline	
\begin{tabular}{c}	
	$\hat{\ell}^{e}_{L}=\begin{pmatrix}\hat{\nu}^{e}\\ \hat{e}^{e} \end{pmatrix}_{L}$  \\
	$\hat{\ell}^{\mu}_{L}=\begin{pmatrix}\hat{\nu}^{\mu}\\ \hat{\mu}^\mu \end{pmatrix}_{L}$  \\
	$\hat{\ell}^{\tau}_{L}=\begin{pmatrix}\hat{\nu}^{\tau}\\ \hat{\tau}^{\tau} \end{pmatrix}_{L}$ 
\end{tabular} &
\begin{tabular}{c}
		$0^{+}$	\\
	\\	$0^{+}$	\\
	\\	$-1^{+}$	\\
\end{tabular} &
\begin{tabular}{c}
		\\
	\\		\\
	\\		\\
\end{tabular}   &
\begin{tabular}{c}
	$ \begin{matrix}\hat{\nu}^{e\; c}_{L}	\\ \hat{\nu}^{\mu\; c}_{L} \end{matrix}$  \\
	$ \begin{matrix}\hat{\nu}^{\tau\; c }_{L}	\\  \hat{e}^{e\; c}_{L} \end{matrix}$  \\
	$ \begin{matrix} \hat{e}^{\mu\; c }_{L}	\\  \hat{e}^{\tau\; c}_{L} \end{matrix}$ 
\end{tabular} &
\begin{tabular}{c}
	$ \begin{matrix} \sfrac{-1}{3}^{-}	\\ \sfrac{-1}{3}^{-} \end{matrix}$  \\
	$ \begin{matrix} \sfrac{-1}{3}^{-}	\\ \sfrac{+4}{3}^{-} \end{matrix}$  \\
	$ \begin{matrix} \sfrac{+1}{3}^{-}	\\ \sfrac{+4}{3}^{-} \end{matrix}$ 
\end{tabular} &
\begin{tabular}{c}
	  \\
	 \\
\end{tabular}
\\ \hline\hline 

\multicolumn{6}{c}{Non-SM Leptons}\\ \hline\hline	
\begin{tabular}{c}	
    $\hat{E}_{L}$	\\
    $\hat{\mathcal{E}}_{L}$	\\
\end{tabular} &
\begin{tabular}{c}
	$-1^{+}$	    	\\
	$\sfrac{-2}{3}^{+}$	\\
\end{tabular} &
\begin{tabular}{c}
			\\
			\\
		
\end{tabular}   &
\begin{tabular}{c}	
    $\hat{E}_{L}^{c}$	\\
    $\hat{\mathcal{E}}_{L}^{c}$	\\
\end{tabular} &
\begin{tabular}{c}
	$\sfrac{+2}{3}^{+}$		\\
	$+1^{+}$	            \\
\end{tabular} &
\begin{tabular}{c}
			\\
			\\
	
\end{tabular}
\\	\hline\hline 

\multicolumn{3}{c}{Majorana Fermions} & 
\begin{tabular}{c}	
	$\mathcal{N}_{R}^{1,2,3}$	\\
\end{tabular} &
\begin{tabular}{c}
	$0^{-}$	\\
\end{tabular} &
\begin{tabular}{c}
	\\
		
\end{tabular}
\\	\hline\hline 
\end{tabular}
\end{table}
Analogously, the lepton superpotencial correspond to the non-SUSY potential; with the fields promoted to superfields and the conjugate Higgs fields promoted to the primed ones. Then the superpotenial reads as follows.

\begin{align}
    W_{L}=&\hat{\ell}_{L}^{p}\hat{\phi_{2}}h_{2\nu}^{pq}(\hat{\nu}_{L}^{q})^{C} +                      \hat{\ell}_{L}^{p}\hat{\phi '_{2}}{h}_{2e}^{p\mu}(\hat{e}_{L}^{\mu})^{C}+ \hat{\ell}_{L}^{\tau}\hat{\phi '_{2}}{h}_{2e}^{\tau r}(\hat{e}_{L}^{r})^{C} + 
        \hat{\ell}_{L}^{p}\hat{\phi '_{1}}{h}_{1E}^{p})\hat{E}_{L}^{C} \nonumber \\
    &+ \hat{E}_{L}\hat{\chi '}{g}_{\chi' R}\hat{E}_{L}^{C} 
        +\hat{E}_{L}\hat{\sigma}h_{\sigma e}^{r}(\hat{e}_{L}^{r})^{C} + \hat{E}_{L}\mu_{E}\mathcal{E}_{L}^{C} + \hat{\mathcal{E}}_{L}\hat{\chi}g_{\chi\mathcal{E}}\hat{\mathcal{E}}_{L}^{C} \nonumber \\
        &+ \hat{\mathcal{E}}_{L}\hat{\sigma '}{h}_{\sigma' e}^{\mu}(\hat{e}_{L}^{\mu})^{C} + \hat{\mathcal{E}}\mu_{\mathcal{E}}\hat{\mathcal{E}}_{L}^{C}+\hat{\nu}_{L}^{jC}\chi ' ({h}_{\chi}^{\prime N})^{ji}N_{L}^{iC}
        + \frac{1}{2}N_{L}^{iC}M_{ij}N_{L}^{jC}, \nonumber\label{WL}
\end{align}

In a similar way, the conditions for implementing a seesaw mechanism are given in the matrix structure. In this case the lightest fermion, identified as the electron, turns out to be massless but it acquire a finite mass value through radiative corrections due to the interaction with the scalar singlet $\sigma$
\begin{align}
    \mathcal{M}_{E}&=\frac{1}{\sqrt{2}}\left(\begin{array}{ c c c |c c}\\
    0                           & h_{2e}^{e\mu}v'_{2}     & 0 &  h_{1e}^{E}v'_{1}    & 0 \\
    0                           & h_{2e}^{\mu\mu}v'_{2}   & 0 &  h_{1\mu}^{E}v'_{1}  & 0 \\
    h_{2e}^{\tau e}v'_{2}  & 0                            & h_{2e}^{\tau\tau}v'_{2} & 0 & 0 \\ \hline
    0 & 0 & 0 & {g}_{\chi' E}v'_{\chi} & -\mu_{E} \\
    0 & 0 & 0 & -\mu_{\mathcal{E}} & g_{\chi\mathcal{E}}v_{\chi}  \\
    \end{array} \right)
\end{align}
\begin{figure}
\centering
\includegraphics[scale=0.7]{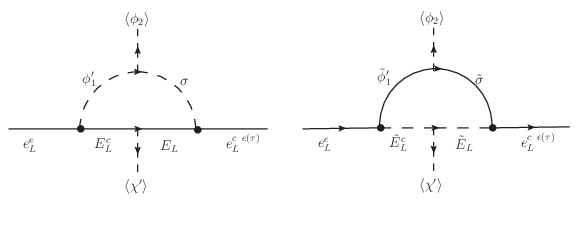}
\caption{Electron's one loop correction to electron's mass.}
\end{figure}

which in this particular case, the SUSY and non-SUSY contributions are given by:
\begin{small}
\begin{align}
    v_{2}\Sigma _{11(13)}^{NS}=\frac{-1}{16\pi ^2}\frac{v_{2}}{\sqrt{2}}\frac{\lambda_{1}\mu_{\sigma}h_{\sigma}^{ec e(\tau)}h_{1E}^{e}}{M_{E}}C_0\left(\frac{m_{h1}^{\prime}}{M_{E}},\frac{m_{\sigma}^{\prime}}{M_E}\right).
\end{align}

\begin{align}
    v_{2}&\Sigma_{11(13)}^{S}(p^{2}=0) = -\frac{1}{32\pi^{2}}\frac{v_{2}}{\sqrt{2}}\sum_{n=1}^{10}\sum_{k=1}^{2}Z_{L}^{\tilde{E}_{L}^{c}n}Z_{L}^{\tilde{E}_{L}n} Z_{L}^{\sigma k}Z_{L}^{\sigma^{\prime} k} \lambda_{1}\mu_{\sigma}h_{\sigma}^{ec e(\tau)}h_{1E}^{e}\times \\
    &\times\left[\frac{(\tilde{m}_{\sigma  k}+\tilde{m}_{h_{1}}^{\prime})^{2}}{\tilde{M}_{L_{n}}^{2}}C_{0}\left(\frac{\tilde{m}_{h1}^{\prime}}{\tilde{M}_{L_{n}}},\frac{\tilde{m}_{\sigma  k}}{\tilde{M}_{L_{n}}} \right) + \tilde{m}_{h1}^{\prime 2}B_{0}(0,\tilde{m}_{\sigma}^{\prime},\tilde{M}_{L_{n}}) + \tilde{m}_{\sigma k}^{2}B_{0}(0,\tilde{m}_{h1}^{\prime},\tilde{M}_{L_{n}}) \right] \nonumber
\end{align}
\end{small}
Anyway, when considering the neutral leptons, the Majorana fermions are expected to be heavy, then the standard model neutrinos acquire small mass values due to seesaw mechanism. The model generates the correct squared mass differences reported for neutrino oscillations \cite{8}. However, that is a result that arise from the no-SUSY version of the model \cite{no-susy}

\section{Conclusions}
 It is shown that the $U(1)_{X}$ extension of the MSSM is compatible with the SM physics and a $125 GeV$ scalar particle through 4 VEV at the electroweak scale. It allows to fermions with different iso-spin to get coupled to different Higgs doublets, and showing that it is possible to get a big top-quark mass since it becomes decoupled and the corresponding VEV is found to have a big value too.  Anyway, it is yet unexplored the neutralino and chargino physics as well as sfermions fenomenology. Furthermore, through parity breaking terms it can be considered the $\mu$-problem, but for it now the model implications are unexplored.

\end{document}